\documentstyle[aps,prl,preprint,psfig,floats]{revtex}
\tightenlines
\begin{document}
\draft
\preprint{UNDQCG-00-01}
\title{Implementation of a quantum phase gate by the optical Kerr effect}
\author{S. Glancy, J. M. LoSecco, C. E. Tanner}
\address{Department of Physics, University of Notre Dame, Notre Dame, Indiana 46556}
\date{\today}
\maketitle
\begin{abstract}
We show that the optical Kerr effect can be used to construct a quantum phase gate. It is well known from quantum nondemolition techniques that, as two photon field modes pass through a Kerr medium, the phase of each mode will be shifted, and the size of the phase shift will depend on the number of photons in both modes. We discuss the Hamiltonian responsible for this effect and show how this can produce an effective photon-photon interaction which corresponds to the quantum phase gate operating on two qubits each of which is represented by the elliptical polarization of one photon field. We discuss decoherence and losses and suggest some methods for dealing with them.\\
\end{abstract}


\narrowtext
\section{The Quantum Phase Gate}
Theoretically quantum computers can perform some types of calculations much faster than classical computers \cite{DeutschFeynman}, but the technological difficulties of manipulating quantum information have so far prevented researchers from constructing a quantum computer which is able to perform useful tasks.  The difficulty of building a quantum computer was greatly diminished when it was realized that a network of quantum phase gates (QPG) operating in the product space of two qubits, single bit rotations, and single bit phase shift gates can constitute a universal quantum computer \cite{BarencoLloyd}. The QPG simply gives the product state of two qubits a phase shift depending on the values of each qubit.  In other words, the QPG performs the operation 
\begin{equation} \left[ \begin{array}{cccc}
1 & 0 & 0 &0 \\
0 & e^{i \alpha} & 0 & 0 \\
0 & 0 & e^{i \beta} & 0 \\
0 & 0 & 0 & e^{i \gamma} \end{array} \right] 
\begin{array}{l}
|00\rangle \\
|01\rangle \\
|10\rangle \\
|11\rangle \end{array}
\end{equation}
in the computational basis of the two qubits.  Provided that $\alpha+\beta \neq \gamma$ (mod $2 \pi$) a network of QPGs supplemented with single bit gates can mimic the operation of any other unitary operator acting on the qubits.

Recently an implementation of a QPG has been demonstrated \cite{harouche} utilizing Rydberg states and a photon in a microwave cavity. In this letter we explain how a QPG that operates in the product space of the polarizations of two photons can be constructed by using the optical Kerr effect.  The photons are made to interact as they pass through a material with a third-order nonlinear susceptibility. These Kerr materials are used for a wide variety of optical applications. In the presence of a superposition of electromagnetic waves at different frequencies and/or in different directions, these materials are used in four-wave mixing applications such as frequency conversion, phase conjugation, real time holography, and image correlation. In the presence of a wave at a single frequency, the refractive index of such materials is intensity dependent and gives rise to the phenomenon of self focusing \cite{photonics}. When a superposition of waves is present, the optical Kerr effect also produces an interaction in which the intensity of one frequency component influences the index of refraction of another frequency component. As described by Mandel and Wolf in \cite{coherence}, this effect can be used to perform quantum non-demolition and back-action evading measurements, during which the intensity of one frequency component can be used to control the phase of another without altering the photon number of either component. Thus without loss of photon number, the frequency components can become entangled in a way that lends itself well to quantum computations.

\section{Elliptical Polarization as a Qubit}
Because a photon's polarization state can be easily manipulated with devices like Faraday rotators and quarter-wave plates, we take our qubit to be a single state
\begin{equation}
|Q\rangle = a|H\rangle +b|V\rangle,
\end{equation}
where $a$ and $b$ are complex numbers \cite{DeutschFeynman}, with
\begin{equation}
|a|^{2}+|b|^{2}=1.
\end{equation}
We implement this as the elliptical polarization of a beam of light, so that the $|H\rangle$ and $|V\rangle$ eigenstates correspond to horizontal and vertical polarized modes of the photon field of a particular frequency. Other all optical qubit implementations have represented qubits with the paths that a laser travels through an interferometer, and the computation is performed using interference. Although it is easy to construct a small quantum computer using this method, the number of optical components (beam splitters and mirrors) increases exponentially with the number of qubits in the computer \cite{Kwait}. Because our proposal encodes the qubits with polarization rather than interferometer paths, it does not suffer from this exponential growth.

The photon characterizing the first qubit $|Q_{1}\rangle$ has frequency $\omega_{1}$, and the second qubit $|Q_{2}\rangle$ has frequency $\omega_{2}$. In order to increase the effectiveness of the nonlinear medium that will carry out the QPG operation, the two photons should travel through it colinearly.  The computer's information is encoded in the photons' polarizations, so the photon frequency is the only means by which we can distinguish the two qubits.

The electric field operator that acts on the photons traveling through the QPG is given by \cite{Imoto}
\begin{equation}
{\bf E}(t, z)=\frac{1}{\sqrt{v}} \sum^{2}_{j=1} \sum_{p=H,V} \sqrt{\frac{\hbar \omega_{j}}{2 \epsilon_{0}}} \left[ i\hat{a}_{jp} \hat{\bf \varepsilon}_{p} e^{-i( \omega_{j}t-k_{j}z)} + \text{h.c.}\right]
\end{equation}
where $t$ is time $z$ is the coordinate describing the direction of propagation, $v$ is the normalization volume, $j$ represents each qubit, $p$ represents the polarization, $\epsilon_{0}$ is the permittivity of free space, $\hat{a}_{jp}$ is the annihilation operator and $\hat{a}_{jp}^{\dagger}$ is the creation operator of a photon with frequency $\omega_{j}$ and polarization $p$, the $\hat{\varepsilon}_{p}$ are unit polarization vectors, and $k_{j}$ is the wave number of qubit $j$.

\section{Optical Kerr Effect}
An optical medium exhibiting the frequency independent Kerr effect is governed by a nonlinear
polarization vector with components 
\begin{equation} P_{l} = \chi E_{l}+  \sum_{m,q,r=1,2,3}\chi^{(3)}_{l,m,q,r} E_{m} E_{q} E_{r}, \end{equation}
where $\chi$ is the linear component of the susceptibility, and $\chi^{(3)}_{l,m,q,r}$ is third order nonlinear response to the electric field, and $l$, $m$, $q$, and $r$, represent the three spatial directions \cite{photonics}. A frequency dependent $\chi^{3}$ cannot be factored so simply but can be expressed through the Fourier transforms of ${\bf P}$ and ${\bf E}$ \cite{coherence,Imoto}.
The behavior of the electric fields representing the two qubits as they travel through the Kerr medium is given by the Hamiltonian
\begin{eqnarray} H = &&\frac{1}{2} \int \frac{1}{\mu_{0}} {\bf B}^2 dv
+ \frac{1}{2} \int \epsilon_{0} {\bf E}^2 dv\nonumber\\
&&+ \frac{1}{2} \int \chi {\bf E}^2 dv
+ \frac{3}{4} \int {\bf E}\cdot{\bf P}^{(3)}_{NL} dv \end{eqnarray}
where $\mu_{0}$ is the permeability of the non-magnetic material, ${\bf B}$ is the total magnetic field, and ${\bf P}^{(3)}_{NL}$ is the nonlinear part of the polarization vector including frequency dependent terms.

\begin{figure}[]
\centerline{\psfig{figure=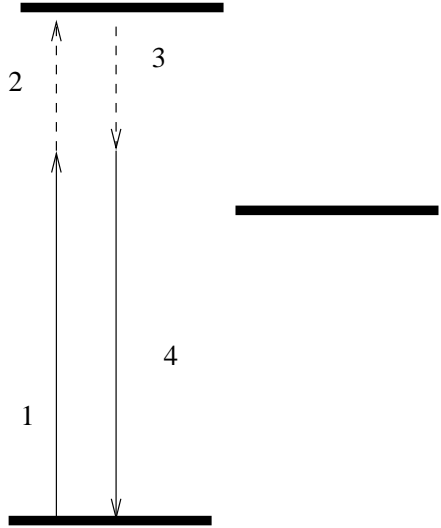}}
\caption{Diagram of a possible transition scheme leading to the Hamiltonian (\ref{Hamiltonian}). The solid arrows represent photons with frequency $\omega_{1}$, and the dashed arrows represent photons with frequency $\omega_{2}$.  The numbers give the order in which each photon creation/annihilation operator appears in (\ref{Hamiltonian}).}
\label{transitions}
\end{figure}

We may simplify this complicated Hamiltonian by choosing the two frequencies of the electric fields to be nearly resonant with excitations of the medium as pictured in Fig.\ \ref{transitions} \cite{Imoto}. Exactly resonant photons will suffer from loss, so it is best that they are slightly detuned. A similar technique, with a different energy transition scheme, has been demonstrated by Sinatra et.\ al.\ in \cite{Sinatra}. In their experiment the two lasers are coupled through a gas of $^{87}\text{Rb}$ where the photons interact with a $\Lambda$-type three-level system. If the photon frequencies are chosen correctly many of the terms in the above Hamiltonian can be ignored. Upon substituting the operator forms of the fields we obtain
\begin{eqnarray}
H = &&\hbar \omega_{1} \frac{\epsilon}{\epsilon_{0}}\left(\hat{n}_{1}+\frac{1}{2}\right)+
\hbar \omega_{2} \frac{\epsilon}{\epsilon_{0}}\left(\hat{n}_{2}+\frac{1}{2}\right)\nonumber\\
&&+ \frac{3\hbar^{2}}{16v\epsilon_{0}^{2}}\omega_{1}\omega_{2}\chi^{(3)}(\omega_{2};-\omega_{1},\omega_{2},\omega_{1})\hat{n}_{1}\hat{n}_{2},
\label{Hamiltonian}
\end{eqnarray}
where $\epsilon=\epsilon_{0}+\chi$ is the linear permittivity of the medium, $\chi^{(3)}(\omega_{2};-\omega_{1},\omega_{2},\omega_{1})$ is the only surviving frequency dependent term in the nonlinear susceptibility, and $\hat{n}_{1}$ and $\hat{n}_2$ are the photon number operators.

As the photons pass through the Kerr medium, they will experience some phase shift, which can be calculated from the time evolution of the annihilation operators $\hat{a}_{1}$ and $\hat{a}_{2}$.  Using Heisenberg's equation of motion, we find that
\begin{equation}
\hat{a}_{1}(t) = e^{-i \omega_{1} t \left(\frac{\epsilon}{\epsilon_{0}} + \tilde{\chi}^{(3)}_{\text{int}} \omega_{2} \hat{n}_{2}\right)} \hat{a}_{1}(0)
\end{equation}
and
\begin{equation}
\hat{a}_{2}(t) = e^{-i \omega_{2} t \left(\frac{\epsilon}{\epsilon_{0}}+ \tilde{\chi}^{(3)}_{\text{int}} \omega_{1} \hat{n}_{1}\right)} \hat{a}_{2}(0),
\end{equation}
where
\begin{equation}
\tilde{\chi}^{(3)}_{\text{int}}=\frac{3\hbar^{2}}{16v\epsilon_{0}}\chi^{(3)}(\omega_{2};-\omega_{1},\omega_{2},\omega_{1})
\end{equation}
\cite{coherence,Imoto}. Therefore each photon field traveling through the Kerr medium will receive a phase shift that depends on the number of photons in the other field.  More specifically, if one field is in a superposition of photon number states, the other field will emerge in an entangled state so that it has been given a superposition of different phase shifts.

\section{Construction of the QPG}
\begin{figure}[t]
\centerline{\psfig{figure=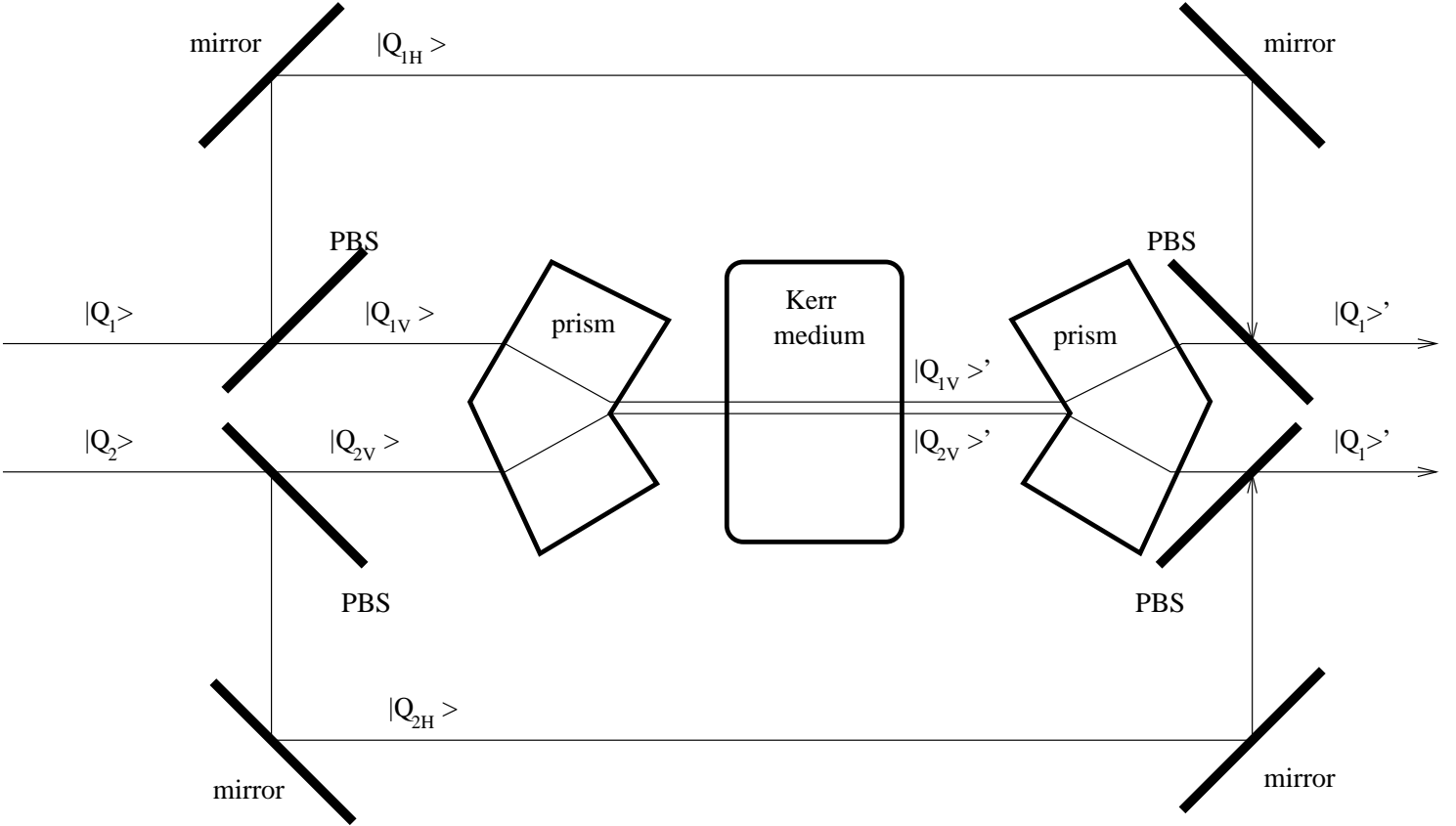}}
\caption{Diagram of proposed QPG. The photons representing $|Q_{1}\rangle$ and $|Q_{2}\rangle$ travel from left to right through the gate.}
\label{device}
\end{figure}

It is relatively easy to manipulate the polarization states of a photon, so we imagine that the quantum computer contains two photons with the states
\begin{equation} |Q_{1}\rangle = a_{1}|H_{1}\rangle + b_{1}|V_{1}\rangle \end{equation}
and
\begin{equation} |Q_{2}\rangle = a_{2}|H_{2}\rangle + b_{2}|V_{2}\rangle . \end{equation}
However the phase shifts provided by the isotropic Kerr effect depend not on polarization, but instead on photon number, so we must translate the above polarization states into Fock states. As depicted in Fig.\ \ref{device}, this can be done by passing each photon through a polarizing beam splitter effectively cutting each qubit into two beams, one with horizontal and one with vertical polarization. The amplitudes for finding zero or one photon in each beam are given by
\begin{equation} |Q_{1H}\rangle = b_{1}|0_{1H}\rangle + a_{1}|1_{1V}\rangle, \end{equation}
\begin{equation} |Q_{1V}\rangle = a_{1}|0_{1V}\rangle + b_{1}|1_{1V}\rangle, \end{equation}
\begin{equation} |Q_{2H}\rangle = b_{2}|0_{2H}\rangle + a_{2}|1_{2H}\rangle, \end{equation}
and
\begin{equation} |Q_{2V}\rangle = a_{2}|0_{2V}\rangle + b_{2}|1_{2V}\rangle. \end{equation}

The two horizontal beams may now be deflected around the Kerr medium while the vertical beams are directed through it.  To achieve the largest nonlinear effects, we combine the two vertical beams with prisms so that they are collinear when entering the medium. Prisms may also be used to separate the beams after they emerge. We assume that the horizontal beams travel for an integer number of wavelengths and receive zero phase shift, while the two vertical beams are phase shifted according to the operators
\begin{equation}
\hat{K}_{1}(t) = e^{-i \omega_{1} t \left(\frac{\epsilon}{\epsilon_{0}} + \tilde{\chi}^{(3)}_{\text{int}} \omega_{2} \hat{n}_{2V}\right)}
\end{equation}
and
\begin{equation}
\hat{K}_{2}(t) = e^{-i \omega_{2} t \left(\frac{\epsilon}{\epsilon_{0}} + \tilde{\chi}^{(3)}_{\text{int}} \omega_{1} \hat{n}_{1V}\right)},
\end{equation}
where $\hat{n}_{1V}$ and $\hat{n}_{2V}$ count the number of photons in $|Q_{1V}\rangle$ and $|Q_{2V}\rangle$.  After emerging from the Kerr medium the vertical parts of the two qubits are
\begin{equation} |Q_{1V}\rangle^{\prime} = a_{1}|0_{1V}\rangle + b_{1} \hat{K}_{1}|1_{1V}\rangle \end{equation}
and
\begin{equation} |Q_{2V}\rangle^{\prime} = a_{2}|0_{2V}\rangle + b_{2} \hat{K}_{2}|1_{2V}\rangle. \end{equation}

The horizontal and vertical parts of the qubits are then recombined at a second pair of polarizing beam splitters, completing the action of the quantum phase gate. If the Kerr medium is anisotropic, it may not be necessary to split the beams into horizontal and vertical parts. Instead the beams would be ``split'' by the medium so that their horizontal and vertical components receive different phase shifts. The output qubits are
\begin{equation} |Q_{1}\rangle^{\prime} = a_{1}|H_{1}\rangle + b_{1} \hat{K}_{1}|V_{1}\rangle \end{equation}
and
\begin{equation} |Q_{2}\rangle^{\prime} = a_{2}|H_{2}\rangle + b_{2} \hat{K}_{2}|V_{2}\rangle. \end{equation}
This is equivalent to the operation of
\begin{equation}
\hat{K}_{1}\otimes\hat{K}_{2} = \left[ \begin{array}{cccc}
1 & 0 & 0 &0 \\
0 & e^{-i \omega_{1} t \frac{\epsilon}{\epsilon_{0}}} & 0 & 0 \\
0 & 0 & e^{-i \omega_{2} t \frac{\epsilon}{\epsilon_{0}}} & 0 \\
0 & 0 & 0 & e^{-it\left(\omega_{1}\frac{\epsilon}{\epsilon_{0}}+\omega_{2}\frac{\epsilon}{\epsilon_{0}}+ 2\omega_{1}\omega_{2}\tilde{\chi}^{(3)}_{\text{int}}\right)} \end{array} \right] 
\begin{array}{l}
|H_{1}H_{2}\rangle \\
|H_{1}V_{2}\rangle \\
|V_{1}H_{2}\rangle \\
|V_{1}V_{2}\rangle \end{array}
\end{equation}
in the product space of $|Q_{1}\rangle \otimes |Q_{2}\rangle$. For correct choices of the parameters $\omega_{1}$, $\omega_{2}$, $t$ (or the length of the Kerr medium), and $\chi^{(3)}(\omega_{2};-\omega_{1},\omega_{2},\omega_{1})$, this is a quantum phase gate satisfying the conditions for universality.

\section{Decoherence and Losses}
Decoherence afflicts all physical implementations of quantum computers, and in this implementation of the QPG, decoherence will mainly come about through the loss of photons. In general the outcome of an interaction with a quantum state is not unique. In most cases the alternate outcomes remove probability from the signal of interest and lead to erratic outcomes when it is measured. Our concept calls for polarization sensitive phase shifts.  But the nonlinear process responsible for the phase shift can have other outcomes too. One needs to have a reasonably large portion of the total interaction rate produce the gated phase shift and to have an ability to reject all the alternate outcomes.  In some experiments, this has been accomplished through the use of single mode fiber \cite{Levenson,White} that only permits the propagation of the, possibly phase shifted, initial wave.  One might also consider the use of filters to remove higher and lower frequencies produced by the nonlinear interaction. As long as the information present in the photons survives the interactions with the gate the device will function properly.

We can imagine that in most calculations, a quantum computer would compute from one polarization eigenstate to another eigenstate. This fact suggests a method for guarding against mistakes due to photon loss. At the output stage of the computer we should have photo-detectors prepared to measure the presence of each photon in both of its $|H\rangle$ and $|V\rangle$ states. Exactly one of the two photo-detectors should register a photon, so if zero or two photons are measured, the calculation should be repeated. Unfortunately the probability of finding an error increases exponentially with the number of qubits that the computer contains. Suppose that $\wp$ is the probability that a single qubit survives all of the computational operations and is correctly detected by its detectors. $\wp$ is likely to be very small for a computer with many qubits and quantum logic gates. If we assume that all $N$ photons must pass through the same number of logic gates and that $\wp$ is not a function of the polarization of the qubits, he probability that all $N$ photons survive the computation and are all detected correctly is then $\wp^{N}$.

The optical Kerr effect depends on the number of photons in the two waves. The induced phase shift is proportional to the intensity of the controlling beam.  Optical losses can reduce the intensity of the beams and change the phase shift associated with the interaction.  One needs a simple way to generate a standard amplitude beam as part of the gate so that the induced phase shift will be a constant. Here we assume that the quantum computer operates at the single photon level, so losses are detected by the absence of an output for the particular qubit.

An intrinsic problem with our approach may be the presence of polarization dependent loses that would make the induced phase shifts depend on the ``state'' of the qubit.  While these may be small, they introduce a form of error into the gate that could accumulate for complex calculations.  It may be possible to compensate for these effects.

In general it may be hard to maintain ridged control on the mechanical structure of the gate to ensure reproduceability.  Without control the gate will populate a small region in the vicinity of the correct operation. We suspect that some sort of feedback mechanism may be needed to insure consistent quality for extended calculations.

Lastly, for most materials the $\chi^{(3)}_{l,m,q,r}$ are known to be very small, and nonlinear effects are observable only for beams with large numbers of photons.  One can combat this by allowing the photons to travel together through long fibers \cite{Levenson}. However, large nonlinear effects exist for photons slightly detuned from resonance transitions in atomic gasses \cite{Sinatra}.

\section{Conclusions}
The photon-photon interaction predicted from the Kerr effect Hamiltonian is the sort of interaction required for the construction of a universal quantum computer. At the single-photon level, this implementation of a QPG is sensitive to loss mechanisms and the requirement of large nonlinear optical effects. Our future work will investigate the extension of this treatment to qubits that are described by large numbers of photons rather than single photons.

\section{Acknowledgments}
We thank James Cushing, Ed Farhi, and Uri Sarid for their helpful discussions. S.~G. thanks the Arthur J.~Schmitt Foundation for fellowship support. J.~M.~L. is supported by the Division of High Energy Physics, Office of Science, U.~S.~Department of Energy under contract number DE-FG02-00ER41145. C.~E.~T. acknowledges support by the Division of Chemical Sciences, Office of Basic Energy Sciences, Office of Science at the U.~S.~Department of Energy under contract number DE-FG02-95ER14579.

\end{document}